\documentclass[11pt,a4paper,english,showpacs,superscriptaddress]{revtex4}

\usepackage{amssymb}
\usepackage{amsmath}
\usepackage{graphicx}
\usepackage{babel}
\usepackage{hyperref}

\usepackage{color}

\begin{document}

\title{Perturbative aspects of mass-deformed ${\cal N}=3$ Chern-Simons-matter theory}

\author{A.~C.~Lehum}
\affiliation{Faculdade de F\'isica, Universidade Federal do Par\'a, 66075-110, Bel\'em, Par\'a, Brazil.}
\email{lehum@ufpa.br}

\author{J. R. Nascimento}
\affiliation{Departamento de F\'{\i}sica, Universidade Federal da Para\'{\i}ba\\
 Caixa Postal 5008, 58051-970, Jo\~ao Pessoa, Para\'{\i}ba, Brazil}
\email{jroberto,petrov@fisica.ufpb.br}

\author{A. Yu. Petrov}
\affiliation{Departamento de F\'{\i}sica, Universidade Federal da Para\'{\i}ba\\
 Caixa Postal 5008, 58051-970, Jo\~ao Pessoa, Para\'{\i}ba, Brazil}
\email{jroberto,petrov@fisica.ufpb.br}

\date{\today}

\begin{abstract}

Within the superfield formalism, we calculate the two-point functions and the effective potential for the mass-deformed ${\cal N}=3$ Chern-Simons-matter theory and discuss the related renormalization group issues.
\end{abstract}

\maketitle

\section{Introduction}

The supersymmetric Chern-Simons theory and related field theory models are intensively discussed within various contexts. The interest to these models was rapidly increased after publishing the seminal paper \cite{MaldAhar} where the ${\cal N}=6$ superconformal Chern-Simons-matter theory has been introduced and intensively discussed within the context of the AdS/CFT correspondence. 
Previous studies on superconformal Chern-Simons theories with various ${\cal N}$ have been carried out in \cite{Schwarz,Gaiotto,Queiruga:2015fzn} and many other papers. However, most of these studies (see f.e. \cite{Bagger,Gromov}) are devoted to the tree-level aspects of these theories. One of a few examples of quantum calculations in Chern-Simons theories with extended supersymmetry has been presented in \cite{BPS} where the one-loop effective action in three-dimensional ${\cal N}=2$ and higher ${\cal N}$-extended supersymmetric theories has been obtained with use of known analogy between ${\cal N}=2$ superspace in $3D$ and ${\cal N}=1$ superspace in $4D$ originally introduced in \cite{Hitchin}. While this correspondence is clearly a powerful tool for study of extended supersymmetric theories (see f.e. \cite{BMS,Gama}),  it is interesting, first, to develop studies of the extended supersymmetric theories with use of ${\cal N}=1$ formalism only, second, to study the models where the higher ${\cal N}$ supersymmetry is broken to ${\cal N}=1$ one. An interesting example of such a theory is the mass-deformed ${\cal N}=3$ Chern-Simons theory \cite{Inba}. In the present paper, we study the quantum aspects of this theory.

The structure of the paper looks like follows. In the section 2, we describe the classical action of the mass-deformed ${\cal N}=3$ Chern-Simons theory. In the section 3, we calculate the two-point functions of scalar and gauge superfields. In the section 4, we consider the one-loop effective potential and find it explicitly for the $SU(N)$ group in large $N$ limit. In the section 5, the renormalization group aspects of the effective potential are presented. The section 6 is a Summary where our results are discussed.

\section{The mass-deformed ${\cal N}=3$ Chern-Simons theory}

Our starting point is the classical action of the mass-deformed ${\cal N}=3$ Chern-Simons theory in ${\cal N}=1$ superspace \cite{Inba}:
\begin{eqnarray}\label{eq1}
S&=&-\int d^5z\Big[-\frac{\kappa}{4\pi}{\rm Tr}\left(-\frac{1}{4}D^{\alpha}\Gamma^{\beta}D_{\beta}\Gamma_{\alpha}+\frac{1}{6}
D^{\alpha}\Gamma^{\beta}\{\Gamma_{\alpha},\Gamma_{\beta}\}+\frac{1}{24}\{\Gamma^{\alpha},\Gamma^{\beta}\}\{\Gamma_{\alpha},\Gamma_{\beta}\}\right)\nonumber\\ 
&-&\frac{1}{2}\sum_{M=1}^2(D^{\alpha}\bar{\Phi}^M+i\bar{\Phi}^M\Gamma^{\alpha})(D_{\alpha}\Phi_M-i\Gamma_{\alpha}\Phi_M)-
m_0(\bar{\Phi}_1\Phi_1-\bar{\Phi}_2\Phi_2)\nonumber\\
&-&\frac{\pi}{\kappa}(\bar{\Phi}_1\Phi_1)(\bar{\Phi}_1\Phi_1)-\frac{\pi}{\kappa}(\bar{\Phi}_2\Phi_2)(\bar{\Phi}_2\Phi_2)+\frac{4\pi}{\kappa}(\bar{\Phi}_1\Phi_1)(\bar{\Phi}_2\Phi_2)+\frac{2\pi}{\kappa}(\bar{\Phi}_1\Phi_2)(\bar{\Phi}_2\Phi_1)
\Big].
\end{eqnarray}
\noindent In the action (\ref{eq1}), the first line presents an usual non-Abelian Chern-Simons action (see f.e. \cite{SGRS}).  We suggest that our gauge fields and ghosts are Lie-algebra valued, $A^{\alpha}=A^{\alpha A}T^A$, $c=c^AT^A$, $c'=c^{\prime A}T^A$, with the generators satisfy the relation ${\rm tr}(T^AT^B)=R\delta^{AB}$.
Our calculations are carried out for an arbitrary gauge group unless we make some special restrictions. Here the fields with the index 1 carry the $SO(2)$ charge $\frac{1}{2}$, and with the index 2 -- the $SO(2)$ charge $-\frac{1}{2}$ (the $\Phi_1$, $\Phi_2$ are the same fields as $\Phi^+$, $\Phi^-$ from \cite{Inba}). We use the notations of \cite{Inba} which are similar to those one from \cite{SGRS}, up to some overall multipliers. We should also add the gauge fixing term 
\begin{equation}
S_{GF}=\frac{\kappa}{16\pi\xi}{\rm Tr}\int d^5z D^{\alpha}\Gamma_{\alpha}D^{\beta}\Gamma_{\beta}
\end{equation}
and the ghost action
\begin{equation}
S_{GH}=\frac{1}{2}{\rm Tr}\int d^5z c' D^{\alpha}(D_{\alpha}+i\Gamma^{\alpha})c,
\end{equation}
which is the same as in the usual ${\cal N}=1$ Chern-Simons theory \cite{SGRS}.

The propagators in our theory can be cast as
\begin{eqnarray}\label{eq4}
\langle \Gamma_{\alpha}(-p,\theta_1)\Gamma_{\beta}(p,\theta_2)\rangle&=&-\frac{\pi i}{\kappa}\frac{D^2}{(p^2)^2}
\left(D_{\beta}D_{\alpha}+\xi D_{\alpha}D_{\beta}\right)\delta_{12}\nonumber\\
&=&-\frac{\pi i}{\kappa}\frac{(1+\xi)C_{\beta\alpha}p^2+(1-\xi)p_{\beta\alpha}D^2}{(p^2)^2}
\delta_{12}~,\nonumber\\
\langle \Phi^{M}(p,\theta_1)\bar{\Phi}^{N}(-p,\theta_2)\rangle&=&i\delta^{MN}\frac{D^2-(-1)^M m_{0}}{p^2+m_0^2}\delta_{12}\nonumber\\
\langle c(p,\theta_1)\bar{c}(-p,\theta_2)\rangle&=&i\frac{D^2}{p^2}\delta_{12}~,
\end{eqnarray}
where $\delta_{12}=\delta^2(\theta_1-\theta_2)$. 

The superficial degree of divergence of our theory is the same as for the usual supersymmetric Chern-Simons-matter theory \cite{CSNC}:
\begin{equation}
\omega=2-\frac{1}{2}(E_{\Gamma}+E_{\Phi})-\frac{N_D}{2},
\end{equation}
where $E_{\Gamma}$ and $E_{\Phi}$ are the numbers of external gauge and scalar legs, respectively, and $N_D$ is the number of spinor supercovariant derivatives acting to external legs. So, the theory is renormalizable. However, this formula does not take into account possibility of mutual cancellation of divergences due to extended supersymmetry, thus, actually the renormalization properties of the theory can be much better.

\section{Two-point functions}

The contributions to the two-point function of the gauge fields arising from the purely gauge/ghost sector can be found along the same lines as in  \cite{CSNC}, and within the one-loop order, even the overall coefficients will not be modified, thus, our results will be just the commutative limit of \cite{CSNC}. The corresponding Feynman diagrams are depicted at the Fig. 1. 

\begin{figure}[htbp]
\includegraphics[scale=0.8]{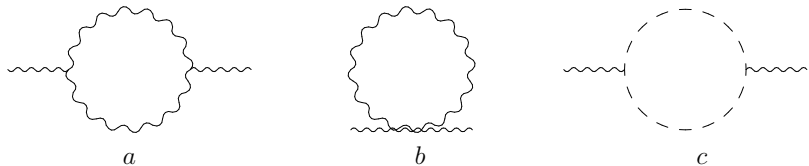}
\caption{Two-point contributions from the gauge sector.}
\end{figure}

Explicitly, we have the following contributions to the effective Lagrangean:
\begin{eqnarray}
\Sigma_a(p)&=&\frac{1}{4R}(1-\xi)\int\frac{d^3k}{(2\pi)^3}\frac{1}{k^2}(N^{ABCC}-2N^{ACBC})A^{\alpha A}A^B_{\alpha} \nonumber\\
\Sigma_b(p)&=&\frac{1}{4R^2}\xi\int\frac{d^3k}{(2\pi)^3}\frac{1}{k^2}(N^{CAD}N^{DBC}-N^{CAD}N^{CBD})A^{\alpha A}A^B_{\alpha} \nonumber\\
\Sigma_c(p)&=& \frac{1}{4R^2}\xi\int\frac{d^3k}{(2\pi)^3}\frac{1}{k^2}(N^{CAD}N^{DBC}-N^{CAD}N^{CBD})A^{\alpha A}A^B_{\alpha},
\end{eqnarray}
where $N^{AB\ldots C}={\rm tr}(T^AT^B\ldots T^C)$. However, while in \cite{CSNC}, it was necessary to consider separately planar and nonplanar parts, and namely the condition of vanishing the IR singularity arising from the non-planar part, implied in restrictions on the gauge group, in our case all contributions are proportional to the same integral $\int\frac{d^3k}{(2\pi)^3}\frac{1}{k^2}$ which identically vanishes within the dimensional regularization. Therefore, the UV divergent one-loop contribution to the two-point function of the gauge field arising from the purely gauge sector is identically zero. Nontrivial superficially divergent contributions to the two-point function of the gauge field can arise only from the coupling to the matter.  The finite parts represent themselves just as the commutative limit of the results found in \cite{CSNC} being equal to
\begin{eqnarray}
\Gamma_{fin}=\int\frac{d^3p}{(2\pi)^3}\int d^2\theta I^{AB}(p)\left((1+\xi^2){\cal L}_{Maxw}^{AB}-2{\cal L}_{GF}^{AB}
\right),
\end{eqnarray}
where ${\cal L}_{Maxw}^{AB}=\frac{1}{2}W_0^{\alpha A}W_{0\alpha}^B$ is  constructed from linearized superfield strengths,  ${\cal L}_{GF}^{AB}=\frac{1}{4}D^{\alpha}\Gamma_{\alpha}^A D^2D^{\beta}\Gamma_{\alpha}^B$ is an analogue of the gauge-fixing term, and $I^{AB}(p)=\int\frac{d^3k}{(2\pi)^3}\frac{1}{k^2(k+p)^2}[N^{CAD}N^{DBC}-N^{CAD}N^{CBD}]$ is the UV finite integral.

Then, the contribution to the two-point function of the gauge field from the matter sector is a natural generalization of the analogous contribution in other three-dimensional supergauge theories. It is generated by two supergraphs depicted at Fig. 2. We note that, unlike \cite{CSNC}, the scalar fields are not Lie-algebra valued in our case being isospinors instead of this.

\begin{figure}[htbp]
\includegraphics[scale=0.8]{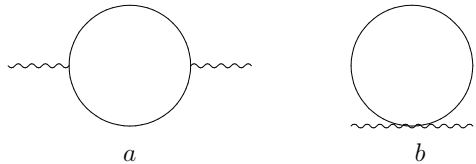}
\caption{Two-point contributions from the scalar sector.}
\end{figure}

Adopting the calculations carried out in \cite{ourqed}, we find that for the (isospinor) scalar field with the mass $m_i$, the two-point function is
\begin{equation}
\label{scal}
\Sigma_{sc,i}(p)=\int d^2\theta f^{AB}_i(p)(W^{A\alpha}_0(-p)W^B_{\alpha 0}(p)+2m_iW^{A\alpha}_0(p)A^B_{\alpha}(p)),
\end{equation}
where $W^{A \alpha}_0$ is the linearized superfield strength, and
\begin{eqnarray}
f^{AB}_i(p)&=&N^{AB}\int \frac{d^3k}{(2\pi)^3} 
\frac{1}{(k^2+m^2_i)[(k+p)^2+m_i^2]}=\nonumber\\
&=&N^{AB}\frac{1}{8\pi|m_i|}+O(p).
\end{eqnarray}
We note that the factor $N^{AB}$ instead of products of two $N^{ABC}$ traces arises since our scalar fields are isospinors under the gauge group rather than Lie-algebra valued (typically, $N^{AB}=R\delta^{AB}$).
Since in our theory (\ref{eq1}) the masses of two scalars are opposite, $m_1=-m_2=m_0$, we find that when two contributions (\ref{scal}) for $\Phi_1$ and $\Phi_2$ are summed, the mass terms are mutually cancelled, and the "kinetic" contributions involving $D^2$ are summed up, hence, for sum of two scalar loops we have
\begin{equation}
\label{scal1}
\Sigma_{sc,total}(p)=\frac{2}{8\pi |m_0|}N^{AB}\int d^2\theta W^{A\alpha}_0W^B_{\alpha 0}+\ldots,
\end{equation}
where dots are for higher-derivative terms. In principle, it is natural to expect that if the contributions of higher orders in the gauge superfields are considered, the full-fledged result will yield the form of the full-fledged super-Yang-Mills action.

So, it remains to consider the contributions with external scalar legs. It is clear that the seagull graph (with one quartic vertex) identically vanishes being proportional to $\int\frac{d^3k}{k^2}$, so, it remains to study the fish contribution. Similarly to \cite{CSNC2}, its divergent part (and hence, a divergent part of the whole two-point function of the scalar superfield) can be shown to vanish in the Landau gauge $\xi=0$. Therefore, in this gauge the two-point function of the scalar identically vanishes.

\section{Effective superpotential at one-loop}

Let us calculate the effective potential in our theory. We can proceed in a manner similar to the ${\cal N}=1$ supersymmetric case \cite{ourCS}. First of all, we note that since the gauge propagator $<\Gamma_{\alpha}\Gamma_{\beta}>$ is proportional to $\left(D_{\beta}D_{\alpha}+\xi D_{\alpha}D_{\beta}\right)$, see (\ref{eq4}), and the $\Gamma_{\alpha}$ field in a triple vertex is accompanied by the $D^{\alpha}$ acting on the adjacent scalar propagator, in the Landau gauge any contribution to the effective potential involving at least one triple vertex, after transporting the $D^{\alpha}$ from the scalar propagator to the gauge one, will yield the factor $D^{\alpha}D_{\beta}D_{\alpha}=0$ and hence vanish (just the same reason implies vanishing of the two-point function of the scalar field). Therefore, there are two types of contributions to the effective potential: the first one is composed by gauge propagators only, with the external scalar fields enter through quartic vertices, and the another one is generated by the scalar sector only, with no gauge propagators.

The first contribution, arising from the gauge-matter couplings, can be calculated in a manner similar to \cite{ourCS}. Indeed, it can be presented as the following trace of the logarithm:
\begin{equation}
\Gamma^{(1)}_{gauge}=\frac{1}{2}\Delta[N]{\rm Tr}\ln(D^{\beta}D^{\alpha}+\frac{1}{\xi}D^{\alpha}D^{\beta}+C^{\alpha\beta}\Sigma^2),
\end{equation}
with $\Delta(N)$ is an algebraic factor arising due to contractions of indices, therefore, taking into account (\ref{eq1}), we find that the final result in the Landau gauge is (cf. \cite{ourCS})
\begin{equation}
\label{gauge0}
\Gamma^{(1)}_{gauge}=-\Delta(N)\frac{\Sigma^4}{64\pi}.
\end{equation}
where $\Phi_M,\bar{\Phi}^M$ are the background matter fields. 
Let us find the factor $\Delta[N]$. It follows from (\ref{eq1}) that the quartic vertex has the explicit form $V_4=\sum\limits_{M=1}^2\int d^5z \bar{\Phi}^{iM}A^{A\alpha}A^B_{\alpha}(T^A)_i^m(T^B)_m^j\Phi_{Mj}$, with $i,j$ are isospinor indices and $A,B$ number the Lie algebra generators. Within the calculation of the one-loop effective potential, Moreover, we suggest that from the all background fields $\Phi^{1i}$, $\bar{\Phi}^{1i}$, only the $a$-th component, i.e. $\Phi_1^a=\Phi_a, \bar{\Phi}^{1a}=\bar{\Phi}_a$, for a some fixed $a$ differs from zero, while the background fields $\Phi_{2i},\bar{\Phi}_{2i}$ are zero for all $i$, i.e. we have spontaneous breaking $SU(N)\to U(1)$, so, our quartic vertex is reduced to $V_4=\int d^5z \bar{\Phi}^{a}A^{A\alpha}A^B_{\alpha}(T^A)_a^m(T^B)_m^a\Phi_{a}$, with no sum over $a$. Since the gauge propagator is proportional to $\delta^{AB}$, we find that the generator-originated algebraic factor from a gauge loop with $n$ vertices is
\begin{eqnarray}
\Delta(N)=(T^A)_a^m(T^B)_m^a(T^B)_a^n(T^C)_n^p\ldots (T^D)_a^q(T^A)_q^a.
\end{eqnarray}
Here we have a sum over indices $m,n,p\ldots q$ while $a$ is fixed. Then, we use the identity for $SU(N)$ generators \cite{Penati}:
\begin{equation}
(T^A)_i^j(T^A)_k^l=2R(\delta_i^l\delta_j^k-\frac{1}{N}\delta_i^k\delta_j^l),
\end{equation}
where ${\rm tr}(T^AT^B)=R\delta^{AB}$ (we can choose $R=\frac{1}{2}$). Hence we have
\begin{eqnarray}
\Delta(N)=(\delta_m^n-\frac{1}{N}\delta_m^a\delta_a^n)(\delta_n^p-\frac{1}{N}\delta_n^a\delta_a^p)\ldots (\delta_s^m-\frac{1}{N}\delta_s^a\delta_a^m).
\end{eqnarray}
The complete result is very involved. So, we find it in the large $N$ limit where all terms proportional to different positive degrees of $1/N$ are suppressed, not only due to the factor $1/N$ but also since there cannot be a possibility to generate any $N$ from contractions because the index $a$ is fixed. The large $N$ result is $\Delta(N)=N+O(1/N)$. At the same time, for our choice of the background we have $\Sigma^2=\bar{\Phi}^{a}\Phi_{a}$. Hence, our result (\ref{gauge0}) for the contribution to the effective potential arisen from the gauge sector, in large $N$ limit, is
\begin{equation}
\label{gauge}
\Gamma^{(1)}_{gauge}=-N\frac{(\bar{\Phi}^{a}\Phi_{a})^2}{64\pi}.
\end{equation}

To calculate the contribution from the purely scalar sector to the effective potential, we impose the same restrictions on the background fields. Again, we choose a particular case $\Phi_2=\bar{\Phi}^2=0$ . As a result, we will have a sum of contributions $\Sigma_1$ formed by cycles of $<\bar{\phi}_{i1}\phi_{j1}>$ propagators and $\Sigma_2$ formed by cycles of $<\bar{\phi}_{i2}\phi_{j2}>$ propagators (remind that our scalars are isospinors). As above, we suggest that from the all isospinors $\Phi_1^i$, $\bar{\Phi}^{1i}$, only the $a$-th component, i.e. $\Phi_1^a, \bar{\Phi}^{1a}$ differs from zero, i.e. we have spontaneous breaking $SU(N)\to U(1)$. In this case, after background-quantum splitting $\Phi^i_{1,2}\to\Phi^i_{1,2}+\phi^i_{1,2}$ our interaction vertex takes the form
\begin{eqnarray}
V_{sc}&=&-\frac{2\pi}{\kappa}(|\Phi_a|^2\bar{\phi}^i_1\phi_{1i})-\frac{\pi}{\kappa}(\Phi_{1a}\Phi_{1a}\bar{\phi}_1^a\bar{\phi}_1^a+\phi_{1a}\phi_{1a}\bar{\phi}_1^a\bar{\phi}_1^a)+\nonumber\\
&+&\frac{4\pi}{\kappa}(|\Phi_a|^2\bar{\phi}^i_2\phi_{12})+\frac{2\pi}{\kappa}(|\Phi_a|^2\bar{\phi}^a_2\phi_{2a}),
\end{eqnarray}
with no sum over $a$. However, since the propagators of scalar superfields are proportional to $\delta^i_j$, we see that the dominant contribution for the large $N$ is contributed by the vertices involving sum over $i$ only, and any vertices involving the fields with a fixed index $a$ will yield only subleading contributions. So, we can suppress these vertices, and our effective potential in large $N$ limit will be given by
\begin{eqnarray}
V^{(1)}_{sc}=N{\rm Tr}\ln (D^2-m_0-\frac{2\pi}{\kappa}|\Phi_a|^2)+N{\rm Tr}\ln (D^2+m_0+\frac{4\pi}{\kappa}|\Phi_a|^2).
\end{eqnarray}
Now we employ the results from \cite{ourEP} and find
\begin{equation}
\label{scalar}
V^{(1)}_{sc}=\frac{N}{16\pi}\left((-m_0-\frac{2\pi}{\kappa}|\Phi_a|^2)^2+(m_0+\frac{4\pi}{\kappa}|\Phi_a|^2)^2
\right).
\end{equation}
The final result of the effective potential is the sum of (\ref{gauge}) and (\ref{scalar}). We see that it involves quadratic and quartic terms.

\section{Renormalization group improvement of the effective superpotential}

In this section we investigate how we can use the Renormalization Group Equation (RGE) in the determination of the effective superpotential. The use of RGE to improve the calculation of the effective potential has been intensively used in non-supersymmetric theories ~\cite{elias:2003zm,Chishtie:2005hr,PhysRevD.72.037902,Meissner:2006zh,Meissner:2008uw,AGQuinto,Chishtie:2010ni,Dias:2010it,Steele:2012av,Chun:2013soa}, and recently the method has been extended to superspace formalism~\cite{Quinto:2014zaa}.

There is no UV divergent renormalization of the coupling constant $\kappa$ nor of the wave-function of the gauge superfield in the ${\cal{N}}=3$ Chern-Simons-matter theory, because the Chern-Simons term only receives up to one-loop corrections~\cite{Blasi:1989mw}, which in three dimensional spacetime are completely regularized. Therefore, the only non-trivial renormalization group function of the model is the anomalous dimension of the matter superfields. 

Since the first UV divergent integrals appear only at two-loop, the anomalous dimension $\gamma_\Phi$ of the matter superfields, just as for the ${\cal N}=2$ Abelian model~\cite{Avdeev:1991za}, in the lowest order has the form
\begin{equation}
\gamma_\Phi=k_1\left(\frac{4\pi}{\kappa}\right)^4 =k_1 g^2,
\end{equation}
\noindent where $g=\left(\frac{4\pi}{\kappa}\right)^2$ and $k_1>0$ is a numerical factor.

The RGE to the effective superpotential in the massless limit is given by
\begin{align}
\left[\mu\frac{\partial}{\partial\mu}+\beta_{g}\frac{\partial}{\partial g}+\gamma_{\Phi}|\Phi_a|\frac{\partial}{\partial|\Phi_a|}\right]V_{eff}\left(|\Phi_a|;\mu,g,L\right) & =0\,,\label{eq:RGE1}
\end{align}
where $\Phi_a$ is the vacuum expectation value of the matter superfields, $\gamma_{\Phi}$
is the anomalous dimension of the scalar matter superfields, $\mu$ is the mass scale introduced by the regularization, and
\begin{align}
L & =\ln\left(\frac{|\Phi_a|^{2}}{\mu}\right).\label{eq:defL}
\end{align}

We shall use for $V_{eff}$ the ansatz
\begin{equation}
V_{eff}=|\Phi_a|^{4}S\left(L\right)\thinspace,\label{eq:KeffAnsatz1}
\end{equation}
where
\begin{equation}
S\left(L\right)=A\left(g\right)+B\left(g\right)L+C\left(g\right)L^{2}+D\left(g\right)L^{3}+\cdots\thinspace,\label{eq:KeffAnsatz2}
\end{equation}
and $A,\thinspace B,\thinspace C,\thinspace D,\thinspace\ldots$ are defined as
series in powers of the coupling constant $g$.

The RGE (\ref{eq:RGE1}) can be conveniently rewriten in terms of $S(L)$. From (\ref{eq:defL}) we can see that $\partial_{L}=\frac{1}{2}|\Phi_a|\partial_{|\Phi_a|}=-\mu\partial_{\mu}$,
and inserting\,(\ref{eq:KeffAnsatz1}) into\,(\ref{eq:RGE1}), we find
\begin{equation}
\left[-\left(1-2\gamma_{\Phi}\right)\partial_{L}+4\gamma_{\Phi}\right]S\left(L\right)=0\thinspace,\label{eq:RGE2}
\end{equation}
where we have used $\beta_g=0$. 

Plugging the ansatz (\ref{eq:KeffAnsatz2}) in (\ref{eq:RGE2}),
and organising the resulting expression by orders of $L$, we obtain a series of equations, of which we write down  the first three:
\begin{equation}
-\left(1-2\gamma_{\Phi}\right)B(g)+4\gamma_{\Phi}A(g)=0\thinspace,\label{eq:orderL0}
\end{equation}
\begin{equation}
-2\left(1-2\gamma_{\phi}\right)C(g)+4\gamma_{\Phi}B(g)=0\thinspace,\label{eq:orderL1}
\end{equation}
and
\begin{equation}
-3\left(1-2\gamma_{\phi}\right)D(g)+4\gamma_{\Phi}C(g)=0\thinspace.\label{eq:orderL1a}
\end{equation}

Since all functions appearing in the above equations are
defined as a series in powers of the coupling constant $g$, we find the following conditions order by order in $g$
\begin{eqnarray}
B^{\left(3\right)}&=&4\gamma_\Phi^{\left(2\right)} A^{\left(1\right)};\label{p1}\\
C^{\left(5\right)}&=&\frac{1}{2}4\gamma_\Phi^{\left(2\right)} B^{\left(3\right)}=\frac{1}{2}\left(4\gamma_\Phi^{\left(2\right)}\right)^2A^{\left(1\right)}\label{p2}\\
D^{\left(7\right)}&=&\frac{1}{3}4\gamma_\Phi^{\left(2\right)} C^{\left(5\right)}=\frac{1}{3!}\left(4\gamma_\Phi^{\left(2\right)}\right)^3A^{\left(1\right)}\label{p3}.
\end{eqnarray}

Writing the effective potential as
\begin{eqnarray}\label{veffsum}
V_{eff}=|\Phi_a|^4 \left(\delta_g+\sum_{n=0}^{\infty} C_n^{ll}g^{2n+1}L^n\right)
\end{eqnarray}
\noindent where $\delta_g$ is the counterterm, we find from Eqs.(\ref{p1},\ref{p2},\ref{p3}) the following recurrence relation
\begin{eqnarray}\label{cll}
C_n^{ll}=\left(\frac{4\gamma_\Phi^{\left(2\right)}}{n~g^2}\right) C_{n-1}^{ll},
\end{eqnarray}
\noindent where we identify $C_0^{ll}=A^{(1)}=1$, $C_1^{ll}=B^{(3)}$, $C_2^{ll}=C^{(5)}$, $C_3^{ll}=D^{(7)}$ and so on.

Inserting (\ref{cll}) into  (\ref{veffsum}) and performing the sum, we obtain the following effective superpotential  
\begin{eqnarray}\label{veffsum01}
V_{eff}= |\Phi_a|^4 \left[\delta_g+g\left(\frac{|\Phi_a|^2}{\mu}\right)^{4 k_1 g^2}\right].
\end{eqnarray}
This effective potential represents all loop order contributions of leading logs.

Imposing the Coleman-Weinberg renormalization condition~\cite{Coleman:1973jx}, 
\begin{eqnarray}
\frac{d^4V_{eff}}{d|\Phi_a|^4}\Big{|}_{|\Phi_a|=\sqrt{\mu}}=4!~g,
\end{eqnarray}
\noindent we find the expression for the counterterm $\delta_g$ as
\begin{eqnarray}\label{cc01}
\delta_g=-\frac{2}{3}\left(256 k_1^4 g^9+320 k_1^3 g^7+140 k_1^2 g^5+25 k_1 g^3\right).\end{eqnarray}

Substituting (\ref{cc01}) into (\ref{veffsum01}), we find
\begin{eqnarray}\label{veffren}
V_{eff}=
-\frac{2}{3} k_1 \left(8 k_1 g^2+5\right) \left(4 k_1 g^2 \left(8 k_1 g^2+5\right)+5\right)g^3 |\Phi_a|^4+g  |\Phi_a|^4 \left(\frac{|\Phi_a|^2}{\mu }\right)^{4k_1 g^2}~.
\end{eqnarray}
Although the model presents a generation of the mass scale $\mu$, there is no spontaneous generation of mass because this effect is incompatible with the perturbative regime, just as in the purely scalar model discussed in~\cite{Coleman:1973jx}.

\section{Summary}

We considered the one-loop effects in the ${\cal N}=3$ supersymmetric Chern-Simons theory.  We explicitly calculated the two-point functions of scalar and gauge fields in the one-loop approximation. It is interesting to note that, in an appropriate gauge, the only nontrivial contribution to the two-point function is the linearized Maxwell term. We calculate also the one-loop effective potential which is a sum of quadratic and quartic terms. However, its explicit form can be found only for large $N$ limit which is known to be an important limit in QCD. 

We also discussed the renormalization group improvements for the effective superpotential. Due to non-trivial anomalous dimension of the matter superfields, we could obtain the complete leading log effective superpotential showing that the model does not exhibit spontaneous generation of mass compatible with perturbation theory. 

The methodology we performed can be naturally applied to other extended supersymmetric Chern-Simons models. In particular, it is natural to expect that it can give some interesting results for ${\cal N}=6,8$ Chern-Simons theories and their noncommutative extensions. We  are planning to perform these studies in our next papers.

{\bf Acknowledgements.} 
This work was partially supported by Conselho Nacional de Desenvolvimento Cient\'{\i}fico e Tecnol\'{o}gico (CNPq). A.C.L. has been partially supported by the CNPq project 307723/2016-0 and 402096/2016-9. The work by A.Yu.P. has been supported by CNPq project 303783/2015-0.

\end{document}